\documentclass{article}
\newcommand{\bfr}{\begin{flushright}}
\newcommand{\efr}{\end{flushright}}
 
\begin{document}
% \eqsec  % uncomment this line to get equations numbered by (sec.num)
\title{Finite Temperature and Density Effects in Higher 
Dimensions with and without Compactifications
%\thanks{Presented at ...}%
% you can use '\\' to break lines
}
\author{Kiyoshi Shiraishi\\
%\address{
Department of Physics, Tokyo Metropolitan University,
Tokyo 158%}
}
\date{Prog. Theor. Phys. {\bf 77} (1987) pp. 1253-1266
}
\maketitle
\begin{abstract}
Expressions for the thermodynamic potential of a Dirac fermion gas are
represented at finite temperature with the chemical potential in an
ultrastatic space $R^d\times S^N$. The high- and low- temperature
expansions for the thermodynamic potential are obtained and, in
particular, strongly degenerate fermi gas is investigated. For the
Candelas-Weinberg model, sufficiently high ``charge'' density prevents
the compactification of the extra space.
\end{abstract}
%\PACS{}

%%%%%%%%%%%%%%%%%%%%%%%%%%%%%%%%%%%%%%%%%%%%%%%%%%%%%%%
\section{Introduction}
%%%%%%%%%%%%%%%%%%%%%%%%%%%%%%%%%%%%%%%%%%%%%%%%%%%%%%%
These days many efforts have been made for pursuits of ``the
unification through higher dimensions.'' It is highly remarkable that
supergravity theories have simple or unique structures in more than
four dimensions.\cite{1}  Recent success in superstring theories
suggests that higher dimensions are required not only for simplicity,
but also the consistency of the theories.\cite{2} Anyway, as far as we
regard the extra dimensions as physical settings, we obviously need to
comprehend the phenomena called ``compactifications'', in order to find
out our four-dimensional world.

Recently, there appear many works investigating the behavior of the
time dependence of the scale factor of extra spaces, that is, the
``Kaluza-Klein cosmology.''\cite{3} Many people intend to explain the
large amount of entropy in our universe in the same scenario. The
scenario is referred to with a slogan ``entropy comes from extra
dimensions''\cite{4} for this attempt. Therefore it is necessary to
calculate the thermodynamic quantities in higher dimensional as well
as curved spaces.

On the other hand, Actor derived the thermodynamic potential in
arbitrary dimensions.\cite{5} Particularly, we are interested in the
case with non-vanishing chemical potentials. It is well known that
symmetry restorations depend on finite density effects as well as
finite temperature effects in four-dimensional universe.\cite{6} Thus,
one can expect that compactifications of the extra spaces are
similarly influenced by finite density and temperature.\cite{7,8}

Since we want to know, at least, about the thermodynamic
quantities before and after the compactifications, we have to
obtain the thermodynamic potentials in curved-spaces. In the present
paper, we derive the thermodynamic potential of a Dirac fermion field in
curved spaces such as $T\times R^d\times S^N$.
 
 The present paper is arranged as follows. In \S 2, we briefly review
the thermodynamics and the effective potential in field theories with
imaginary time formalism.\cite{9} We treat only with a Dirac field
throughout this paper. In \S 3, we show the high temperature expansion
of the thermodynamic potentials in the space whose background
metrics are that of $T\times R^d\times S^N$. The low temperature case
is studied in \S 4. The strong]y degenerate femi gas in
higher-dimensional and/or curved spaces are studied. In \S 5, we discuss
the ``Kaluza-Klein thermodynamics,''\cite{10} especially in the case
of non-vanishing chemical potentials. The last section is devoted to a
summary and discussion.

%%%%%%%%%%%%%%%%%%%%%%%%%%%%%%%%%%%%%%%%%%%%%%%%%%%%%%%
\section{The thermodynamic potential and the path integral}
%%%%%%%%%%%%%%%%%%%%%%%%%%%%%%%%%%%%%%%%%%%%%%%%%%%%%%%
We introduce the thermodynamic potential $\Omega$ for an ensemble at
temperature $\beta^{-1}$ and chemical potential $\mu$ as follows:
\begin{equation}
Z_G=\exp(-\beta\Omega)={\rm Tr~} \exp\{-\beta(\hat{H}-\mu\hat{N})\}\,.
\label{2.1}
\end{equation}
Here $\hat{H}$ is the Hamiltonian of the system and $\hat{N}$ is the
particle number operator. $Z_G$ is the so-called grand partition
function. Equation (\ref{2.1}) reads for fermions, after the mode
expansion as
\begin{eqnarray}
Z_G&=&\exp(-\beta\Omega)\nonumber \\
   &=&\prod_k[\exp(-\beta\omega_k)(1+\exp(-\beta(\omega_k-\mu)))(1
+\exp(-\beta(\omega_k+\mu)))]\,,
\label{2.2}
\end{eqnarray}
where the frequency of modes is given by $\omega_k=({\bf
k}^2+M^2)^{1/2}$. Here the zero-point oscillation is included.

Now the entropy $S$, the pressure $P$ and the particle number ${\cal N}$
for the system under consideration are given by the partial derivatives
of $\Omega(\beta, V, \mu)$,
\begin{equation}
S=\beta^2\frac{\partial\Omega}{\partial\beta}\,;\quad
P=-\frac{\partial\Omega}{\partial V}\,;\quad
{\cal N}=-\frac{\partial\Omega}{\partial\mu}\,.       
\label{2.3}
\end{equation}

The ``particle number'' introduced here represents ``the particle
number minus antiparticle number," which may be called ``the charge
asymmetry''.

In the path integral language, the thermodynamic potential for a
Dirac field with mass $M$ is related to the one-loop effective
potential,\cite{5}
\begin{equation}
\ln{\rm Det} [D\!\!\!\!/+M]
   ={\rm Tr} \ln [(\omega_n+i\mu)^2+\omega_k^2]\quad
\mbox{with}\quad
\omega_k^2={\bf k}^2+M^2\,,
\label{2.4}
\end{equation}
where ${\rm Tr}$ means the integration and summation of all physical
modes and degrees of freedom (i.e., including a trace of the Dirac
matrix). Here, $\omega_n=\{2\pi/\beta\}(n+1/2)$, ($n$ is an
integer) is introduced in the imaginary time formalism for field
theories at finite temperature.\cite{9} The chemical potential in
(\ref{2.4}) can be regarded as the zeroth component of (imaginary) gauge
field which is coupled with the charge density of the
field.\cite{5,11}

We can find the relation between one-loop effective potential given
by (\ref{2.4}) and the ``quantum mechanical''\cite{8} expression of the
thermodynamic potential (see (\ref{2.2})) through the following
identities (for fermions):\cite{12}
\begin{eqnarray}
&&\sum_n\ln\left[\left(\frac{2\pi}{\beta}(n+1/2)+i\mu
\right)^2+y^2\right]\nonumber \\
&&=\beta y+ \ln(1+\exp\beta(\mu-y))+\ln(1+\exp-\beta(\mu+y))\,.       
\label{2.5}
\end{eqnarray}

It is well known that the quantum vacuum energy at zero temperature
comes from the first term on the right-hand side of (\ref{2.5}). Apart
from this zero-point energy, we derive the thermodynamic potential for a
Dirac field in flat $d$-dimensional space using (\ref{2.5}) (and
expansions of the logarithms):\cite{5}
\begin{eqnarray}
\Omega&=&({\rm tr}\,{\bf 1})
\frac{V_d}{(4\pi)^{(d+1)/2}}\beta^{-(d+1)}\sum
_{n=1}^\infty(-1)^n\cosh(\beta\mu n)\nonumber \\
&&\times 2\left(\frac{2\beta M}{n}\right)^{(d+1)/2}K_{(d+1)/2}(\beta Mn)
\label{2.6}
\end{eqnarray}
where $K_\nu(x)$ denotes the modified Bessel function, $V_d$ is the
volume of $d$-dimensional space and ${\rm tr}{\bf 1} =2^{[(d+1)/2]}$.
 
We know another way of evaluating the one-loop effective potentials. In
general, the one-loop quantum corrections contain divergences which
need to be regularized. Using the zeta function
regularization,\cite{13} we obtain the following expressions for
effective potentials:
\[
\beta\Omega=\zeta'(0)+\ln(2\pi\mu_R^2)\zeta(0) \,,
\]
where
\begin{eqnarray}
\zeta(s)&\equiv& \frac{{\rm tr}\,{\bf 1}}{2}\frac{V_d}{\Gamma(s)}
\int_0^\infty dt\,t^{s-1}\int\frac{d^d{\bf k}}{(2\pi)^d}\nonumber \\
&&\times
\sum_{n=-\infty}^\infty\exp\left[-t\left\{\left(\frac{2\pi}{\beta}
\left(n+\frac{1}{2}\right)+i\mu\right)^2+{\bf
k}^2+M^2\right\}\right]\, ,
\label{2.7}
\end{eqnarray}
where $\mu_R^2$ is a parameter which has the dimension of mass and comes
from adjusting the scale of the measure of the path integral.\cite{13}
Very recently, Allen showed that $\mu_R^2$ is only involved in the
vacuum energy part of the thermodynamic potential which is independent
of $\beta$.\cite{14} Therefore, as far as we discard the vacuum energy,
we do not need to worry about the regularization-scale $\mu_R^2$ and the
thermodynamic potential of the system can be derived from the first
term of (\ref{2.7}). To see this, we use the following identity (see the
Appendix):
\begin{eqnarray}
&&\sum_{n=-\infty}^\infty\exp\left\{-t\left(\frac{2\pi}{\beta}
\left(n+\frac{1}{2}\right)+i\mu\right)^2\right\}\nonumber \\
&&\qquad 
=\frac{\beta}{(4\pi)^{1/2}}t^{-1/2}\left[1+2\sum_{n=1}^\infty
(-1)^n\cosh(n\beta\mu)\exp\left(
-\frac{\beta^2 n^2}{4t}\right)\right]\,.
\label{2.8}
\end{eqnarray}
Then we can divide $\zeta(s)$ into two parts:
\begin{eqnarray}
\zeta(s)&=&\zeta_0(s)+\zeta_\beta(s) \,,\nonumber \\
\zeta_0(s)&=&\frac{{\rm tr}\,{\bf
1}}{2}\frac{\beta}{(4\pi)^{1/2}}\frac{V_d}{\Gamma(s)}
\int_0^\infty dt\,t^{s-3/2}\int\frac{d^d{\bf k}}{(2\pi)^d}
\exp\left\{-t\left({\bf
k}^2+M^2\right)\right\}\, ,\nonumber \\
\zeta_\beta(s)&=&\frac{{\rm tr}\,{\bf
1}}{2}\frac{\beta}{(4\pi)^{1/2}}\frac{V_d}{\Gamma(s)}
\int_0^\infty dt\,t^{s-3/2}\int\frac{d^d{\bf k}}{(2\pi)^d}
2\sum_{n=1}^\infty(-1)^n\cosh(n\beta\mu)\nonumber \\
&&\times\exp\left\{-t\left({\bf
k}^2+M^2\right)
-\frac{\beta^2 n^2}{4t}\right\}\, .
\label{2.9}
\end{eqnarray}

$\zeta'_0(0)$ contributes to the thermodynamic potential as the
temperature-independent vacuum energy. Thus, we only consider the
contribution from $\zeta'_\beta(0)$. In fact, $\Gamma(s)\zeta_\beta(s)$
does not diverge when $s\rightarrow 0$ and $\Gamma(s)\sim s^{-1}$, so we
can easily find:
\begin{eqnarray}
\Omega&=&\frac{1}{\beta}\zeta'_\beta(0)\nonumber \\
&=&\frac{{\rm tr}\,{\bf
1}}{2}\frac{V_d}{(4\pi)^{1/2}}
\int_0^\infty dt\,t^{-3/2}\int\frac{d^d{\bf k}}{(2\pi)^d}\nonumber \\
&&\times2\sum_{n=1}^\infty(-1)^n\cosh(n\beta\mu)
\exp\left\{-t\left({\bf
k}^2+M^2\right)
-\frac{\beta^2 n^2}{4t}\right\}\nonumber \\
&=&\frac{{\rm tr}\,{\bf
1}}{2}\frac{V_d}{(4\pi)^{(d+1)/2}}
\int_0^\infty dt\,t^{-(d+1)/2-1}\nonumber
\\ &&\times 2\sum_{n=0}^\infty(-1)^n\cosh(n\beta\mu)
\exp\left\{-tM^2
-\frac{\beta^2 n^2}{4t}\right\}\nonumber \\
&=&{{\rm tr}\,{\bf
1}}\frac{V_d}{(4\pi)^{(d+1)/2}}
\beta^{-(d+1)} \sum_{n=1}^\infty(-1)^n\cosh(n\beta\mu)\nonumber
\\&&\times
2\left(\frac{2\beta M}{n}\right)^{(d+1)/2}K_{(d+1)/2}(\beta Mn)\,.
\label{2.10}
\end{eqnarray}
This coincides with (\ref{2.6}).

%%%%%%%%%%%%%%%%%%%%%%%%%%%%%%%%%%%%%%%%%%%%%%%%%%%%%%%
\section{The high-temperature expansion}
%%%%%%%%%%%%%%%%%%%%%%%%%%%%%%%%%%%%%%%%%%%%%%%%%%%%%%%
In this section, let us derive the thermodynamic potential in curved
space. We consider the static backgrouud metric of $R^d \times S^N$ as a
simple but sufficiently general case. In order to treat this case, we
generalize (\ref{2.10}) as follows:
\begin{eqnarray}
\Omega&=&\frac{{\rm tr}\,{\bf
1}}{2}\frac{V_d}{(4\pi)^{(d+1)/2}}
\int_0^\infty dt\,t^{-(d+1)/2-1}2\sum_{n=1}^\infty(-1)^n\cosh(n\beta\mu)
\nonumber
\\ &&\times \sum_{l=0}^\infty (2 d_l)
\exp\left\{-t(\omega_l^2+M^2)
-\frac{\beta^2 n^2}{4t}\right\}\,,       
\label{3.1}
\end{eqnarray}
where
\[
d_l=\frac{\Gamma(N+l)}{l!\Gamma(N)}\,,\quad
\omega_l=\left(l+\frac{N}{2}\right)\frac{1}{a}
\]
and ${\rm tr}{\bf 1}=2^{[D/2]}$ with $D=d+N+1$. $a$ is the radius of the
hypersphere
$S^N$.

 Now, we consider high-temperature case such as $\beta/a\ll 1$. In this
case, we can use the following formula for infinite sum related with
theta function to expand (\ref{3.1}) with respect to $\beta/a$:
\begin{eqnarray}
&&\sum_{l=0}^\infty
d_l\exp\left\{-\left(l+\frac{N}{2}\right)^2
x\right\}\nonumber \\
&&\qquad=\frac{1}{2}\frac{\Gamma(N/2)}{\Gamma(N)}x^{N/2}
\left[1-\frac{1}{12}N(N-1) x+O(x^2)\right]\,. 
\label{3.2}
\end{eqnarray}

This identity is explained in terms of theta function. (See the
Appendix.) Making use of Eq. (\ref{3.2}), we can expand Eq. (\ref{3.1})
as
\begin{eqnarray}
\Omega&=&({\rm tr}\,{\bf
1})\frac{V_dV_N}{(4\pi)^{(d+N+1)/2}}
\beta^{-(d+N+1)} \sum_{n=1}^\infty(-1)^n\cosh(n\beta\mu)\nonumber
\\&&\times
\left[2\left(\frac{2\beta M}{n}\right)^{(d+N+1)/2}K_{(d+N+1)/2}(\beta
Mn)\right.\nonumber \\
&&\quad\left.-2\frac{N(N-1)}{12}\frac{\beta^2}{a^2}\left(\frac{2\beta
M}{n}\right)^{(d+N-1)/2}K_{(d+N-1)/2}(\beta Mn)+\cdots\right]\,,
\label{3.3}
\end{eqnarray}
where $V_N=2\pi^{(N+1)/2}a^N/\Gamma((N+1)/2)$ is the volume of $S^N$.
This first term on r.h.s. of Eq. (\ref{3.3}) is the thermodynamic
potential in flat
$(d+N)$-dimensional space, while the second and further terms
correspond to the deviation from the flat-space case.

When $\mu=0$, one can and the same expression as Dowker's.\cite{15} For
even $D(=1+ d +N)$, the summation over $n$ becomes a polynomial (i.e.,
finite terms).\cite{5} We show here the expression in the simple
massless case for later convenience:
\begin{eqnarray}
\Omega&=&({\rm tr}\,{\bf
1})\frac{V_dV_N}{(4\pi)^{(d+N+1)/2}}
\beta^{-(d+N+1)} \nonumber
\\&\times&
\left[2^{d+N+1}\Gamma\left(\frac{d+N+1}{2}\right)\sum_{n=1}^\infty(-1)^n
\frac{\cosh(n\beta\mu)}{n^{d+N+1}}\right.\nonumber
\\
&&\left.-2^{d+N-1}\Gamma\left(\frac{d+N-1}{2}\right)
\frac{N(N-1)}{12}\frac{\beta^2}{a^2}\sum_{n=1}^\infty(-1)^n
\frac{\cosh(n\beta\mu)}{n^{d+N-1}}+\cdots\right]\,.
\label{3.4}
\end{eqnarray}

%%%%%%%%%%%%%%%%%%%%%%%%%%%%%%%%%%%%%%%%%%%%%%%%%%%%%%%
\section{The low temperature expansion and the strongly degenerate
Fermi gas}
%%%%%%%%%%%%%%%%%%%%%%%%%%%%%%%%%%%%%%%%%%%%%%%%%%%%%%%
Actor gave the low temperature expansion of the thermodynamic
potential.\cite{5} However, the low-temperature approximation becomes
meaningless when $M^2< \mu^2$. When $M^2< \mu^2$, it is well known that
Femi gas degenerates strongly at low temperature. Fortunately, we know
the formula for the polylogarithmic functions ${\rm Li}_N(x)
=\sum_{n=1}^\infty x^n/n^N$:\cite{16}
\begin{eqnarray}
{\rm Li}(-y^{-1})&=&(-1)^{N-1}{\rm Li}_N(-y)\nonumber \\
&&   -\sum_{r=0}^{N-2}\frac{(-1)^r}{r!}(\ln
y)^r[1-(-1)^{N-1-r}](1-2^{r-N+1})\zeta(N- r)\nonumber \\
&&    -\frac{(-1)^N}{N!}(\ln y)^N\, ,          
\label{4.1}
\end{eqnarray}
and this gives another method for expanding $\Omega$ at low temperature.
Hereafter, we consider $M=0$ case, for simplicity. The generalization
to the massive case is straightforward.

In $d$-dimensional flat space (which means dimension of space-time is $1
+ d$), the thermodynamic potential for a massless Dirac field is given
by (cf. (\ref{3.4}))
\begin{eqnarray}
&&\Omega=({\rm tr}{\bf 1})\frac{V_d}{(4\pi)^{(d+1)/2}}
\beta^{-(d+1)}2^{d+1}\Gamma\left(\frac{d+1}{2}
\right)\sum_{n=1}^\infty (-1)^n
\frac{\cosh(n\beta\mu)}{n^{d+1}}\nonumber \\
&=&\left(\frac{{\rm tr}{\bf 1}}{2}\right)\frac{V_d}{(4\pi)^{(d+1)/2}}
\beta^{-(d+1)}2^{d+1}\Gamma\left(\frac{d+1}{2}\right)
[{\rm Li}_{d+1}(-e^{\beta\mu})+{\rm Li}_{d+1}(-e^{-\beta\mu})]\,.
\label{4.2}
\end{eqnarray}
Applying (\ref{4.1}) to (\ref{4.2}), we can reexpress it as
\begin{eqnarray}
\Omega&=&\frac{{\rm tr}{\bf
1}}{2}\frac{V_d}{(4\pi)^{(d+1)/2}}
\beta^{-(d+1)}2^{d+1}\Gamma\left(\frac{d+1}{2}\right)\nonumber \\
&&\times
\left[\frac{-(\beta\mu)^{d+1}}{(d+1)!}\left\{1+(d+1)!
\sum_{b=1}^{[(d+1)/2]}2\zeta(2b)\frac{(\beta\mu)^{-2b}}{(d+1-2b)!}
(1-2^{1-2b})\right\}\right.\nonumber \\
&&\qquad\left.+\{1+(-1)^d\}{\rm
Li}_{d+1}(-e^{-\beta\mu})
\right]\,.
\label{4.3}
\end{eqnarray}
It is apparent that this expression enables us to approximate itself
at low temperature by evaluating the summation and the polylogarithmic
functions appropriately.

Furthermore, if $d+1$ is even, we get the exact form of the
thermodynamic potential which is expressed as polynomials:
\begin{eqnarray}
\Omega&=&-\frac{{\rm tr}{\bf
1}}{2}\frac{V_d}{(4\pi)^{d/2}}\frac{1}{\Gamma\left(\frac{d+2}{2}\right)}
\frac{\mu^{d+1}}{d+1}\nonumber \\
&&\times
\left[1+(d+1)!\sum_{b=1}^{(d+1)/2}\frac{(\beta\mu)^{-2b}}{(d+1-2b)!}
2(1-2^{1-2b})\zeta(2b)\right]\nonumber \\
&=&-\frac{{\rm tr}{\bf
1}}{2}\frac{V_d}{(4\pi)^{(d+1)/2}}\beta^{-(d+1)}2^{d+1}
{\Gamma\left(\frac{d+1}{2}\right)}\nonumber \\
&&\times
\left[\sum_{c=0}^{(d-1)/2}\frac{(\beta\mu)^{2c}}{(2c)!}
2\zeta(d+1-2c)
(1-2^{2c-d})+\frac{(\beta\mu)^{d+1}}{(d+1)!}\right]\,.
\label{4.4}
\end{eqnarray}
This is exactly the same expression as the one obtained by the
high-temperature expansion.\cite{5}

Let us turn our attention to the space $R^d\times S^N$ again.
Integration over $t$ in (\ref{3.1}) yields when $M=0$, (cf.
(\ref{2.10}))
\begin{eqnarray}
\Omega&=&({\rm tr}{\bf 1})\frac{V_d}{(4\pi)^{(d+1)/2}}\beta^{-(d+1)}
\sum_{n=1}^\infty (-1)^n
\cosh(\beta\mu n)\nonumber \\
&&\times
\sum_{l=0}^\infty
(2d_l)2\left(\frac{2\beta\omega_l}{n}\right)^{(d+1)/2}
K_{(d+1)/2}(\beta\omega_l n)\,.
\label{4.5}
\end{eqnarray}
Using the integral representation
\begin{equation}
K_\nu(z)=\frac{\sqrt{\pi}(z/2)^\nu}{\Gamma\left(
\nu+\frac{1}{2}\right)}\int_1^\infty
e^{-zx}(x^2-1)^{\nu-1/2}dx\,,
\label{4.6}
\end{equation}
the sum over $n$ can be performed to give
\begin{eqnarray}
\Omega&=&-({\rm tr}{\bf
1})\frac{V_d}{(4\pi)^{d/2}}\frac{1}{\Gamma
\left(\frac{d+2}{2}\right)}\sum_l d_l\omega_l^{d+1}\nonumber \\
&&\quad\times\int_1^\infty
dx\,(x^2-1)^{d/2}\left\{\frac{1}{e^{\beta(\omega_l
x-\mu)+1}}+(\mu\rightarrow -\mu)
\right\}\,.
\label{4.7}
\end{eqnarray}
Now we can analyze the low-temperature case by using the method
which can be found in many textbooks.\cite{17}

First of all, we examine the zero-temperature $(\beta\rightarrow\infty)$
case. Note, in the limit $\beta\rightarrow\infty$,
\begin{equation}
\frac{1}{e^{\beta
x}+1}\stackrel{\beta\rightarrow\infty}{\longrightarrow}\theta(-x)\,
,           
\label{4.8}
\end{equation}
where $\theta(x)$ is the step function. The expression (\ref{4.7})
reduces to the following form by means of (\ref{4.8}) at zero
temperature:
\begin{equation}
\Omega=-{\rm tr}{\bf
1}\frac{V_d}{(4\pi)^{d/2}}\frac{1}{\Gamma
\left(\frac{d+2}{2}\right)}\sum_{l=0}^{l_m} d_l\omega_l^{d+1}
\int_1^{\mu/\omega_l}
(x^2-1)^{d/2}dx\,,
\label{4.9}
\end{equation}
where $l_m$ is the largest integer satisfying $\omega_{l_m}<\mu$ and
if $\omega_0>\mu$, then $\Omega=0$.
 
As a check on the efficiency of this approach, let us consider the
limit $a\rightarrow\infty$. In this situation, the sum over $l$ is
reduced to the integration by the substitutions:
\begin{equation}
\omega_l\rightarrow z\,,\quad
\sum_{l=0}^{\l_m}d_l\rightarrow\frac{a^N}{\Gamma(N)}\int_0^\mu
dz\,z^{N-1}\,.
\label{4.10}
\end{equation}
Consequently, we can show:
\begin{eqnarray}
\Omega&\sim&-{\rm tr}{\bf
1}\frac{V_d}{(4\pi)^{d/2}}\frac{1}{\Gamma
\left(\frac{d+2}{2}\right)}\frac{a^N}{\Gamma(N)}
\int_0^{\mu}dz\,z^{N+d}\int_1^{\mu/z}dx\,
(x^2-1)^{d/2}\nonumber \\
&=&-{\rm tr}{\bf
1}\frac{V_dV_N}{(4\pi)^{(d+N)/2}}\frac{1}{\Gamma
\left(\frac{d+2}{2}\right)\Gamma
\left(\frac{N}{2}\right)}\frac{\mu^{d+N+1}}{d+N+1}
\int_0^1dy\,y^{N-1}
(1-y^2)^{d/2}\nonumber \\
&=&-\frac{{\rm tr}{\bf
1}}{2}\frac{V_dV_N}{(4\pi)^{(d+N)/2}}\frac{1}{\Gamma
\left(\frac{d+N+2}{2}\right)}\frac{\mu^{d+N+1}}{d+N+1}\,.
\label{4.11}
\end{eqnarray}
As is expected, this result corresponds to the one obtained from
(\ref{4.3}) in flat $(d+N)$-dimensional space at zero temperature.

Particularly, we are interested in the case $d=3$. In this case, one
finds:
\begin{eqnarray}
\Omega&=&-4\frac{V_3}{24\pi^2}\sum_{l=0}^{l_m}d_l
\left[\mu(\mu^2-\omega_l^2)^{1/2}\left(\mu^2-\frac{5}{2}
\omega_l^2\right)\right.\nonumber \\
&&\qquad\qquad\qquad\qquad
\left.+\frac{3}{2}\omega_l^4\ln\left\{\frac{\mu}{\omega_l}+\left(
\frac{\mu^2}{\omega_l^2}-1\right)^{1/2}\right\}\right]\,.
\label{4.12}
\end{eqnarray}
From this expression, we realize that (\ref{4.12}) is merely the sum of
the thermodynamic potential in four-dimensional space-time at zero
temperature \cite{18} over the discrete mass levels which arise from
compactification of $N$ dimensions. The higher-modes than the chemical
potential are frozen out. This interpretation is easily understood by
drawing the figure which is analogous to the one in the paper by Barr
and Brown,\cite{4} but where we replace $T$ with $\mu$. Of course, in
our case, the extra space $S^N$ does not admit zero modes, therefore the
situation is slightly different. If we consider the case that the
extra-space has zero modes, we will obtain the exact expression for
$\Omega$ in the flat
$d$-dimensiollal space when $\mu$ is smaller than the first non-zero
massive mode.

Before considering finite but still low temperature effect we show
another example, the thermodynamic potential in $S^N$(i.e., $d=0$),
\begin{eqnarray}
\Omega&=&-({\rm tr}{\bf 1})\sum_{l=0}^{l_m}d_l(\mu-\omega_l)\nonumber \\
&=&-({\rm tr}{\bf 1})\frac{(l_m+N)!}{l_m!(N-1)!}
\left(\frac{\mu a}{N}-\frac{l_m+(N+1)/2}{N+1}\right)
\frac{1}{a}
\,.
\label{4.13}
\end{eqnarray}
where
\[
l_m=\left[\mu a-\frac{N}{2}\right]\,.
\]
By using relationship (\ref{2.3}), we derive the particle number
\begin{equation}
{\cal
N}=-\frac{\partial\Omega}{\partial\mu}=({\rm
tr}{\bf 1})\frac{(l_m+N)!}{l_m!N!}\,.         
\label{4.14}
\end{equation}
We can see the effect of the discreteness of the Kaluza-Klein mass
level from (\ref{4.14}). One can also find self-consistent solutions of
Einstein equations after balancing between the stress tensor of the
degenerate Fermi gas with the Casimir effect and the cosmological
constant as the case at finite temperature considered by Dowker et
al.\cite{19} The cosmological constant in this case plays an
analogous role to the bag constant in MIT bag model.\cite{18} Although
the above example may be an interesting exercise, we do not study this
situation in this paper.
 
Now, let us return to treating low-temperature effect applying the
textbook-formula \cite{17} to (\ref{4.7}), we can get the following
low-temperature expansion for $\Omega$:
\begin{eqnarray}
\Omega=&-&{\rm tr}{\bf
1}\frac{V_d}{(4\pi)^{d/2}}\frac{1}{\Gamma
\left(\frac{d+2}{2}\right)}\sum_{l=0}^{l_m} d_l\left[
\omega_l^{d+1}
\int_1^{\mu/\omega_l}dy\,
(y^2-1)^{d/2}\right.\nonumber \\
&&+\frac{1}{\beta^2}\frac{\pi^2}{6}d\cdot \mu (\mu^2-\omega_l^2)^{d/2-1}
+\frac{1}{\beta^4}\frac{7\pi^4}{3\times 5!}d(d-2)\nonumber \\
&&\left.\times\mu\{(d-1)\mu^2-3\omega_l^2\}(\mu^2-\omega_l^2)^{d/2-3}
+\cdots\right]\,.
\label{4.15}
\end{eqnarray}

For the manifold which has zero-modes in contrast with $S^N$ of our
case, one can find that the expression for the thermodynamic potential
agrees with that in flat $d$-dimensional space derived by (\ref{4.3})
when
$0<\mu<$ (the mass of the lowest massive mode).

%%%%%%%%%%%%%%%%%%%%%%%%%%%%%%%%%%%%%%%%%%%%%%%%%%%%%%%
\section{Kaluza-Klein thermodynamics and instability at finite density}
%%%%%%%%%%%%%%%%%%%%%%%%%%%%%%%%%%%%%%%%%%%%%%%%%%%%%%%
In this section, We first examine the themodynamics in the space
$R^d\times S^N$. Simllar case with vanishing chemical potential is
extensively investigated by Tosa,\cite{10} so we will concentrate our
attention to the case at finite density. The remainder of this section
is devoted to the investigation of instabilities of compactifications
induced by quantum effects \cite{20} at finite density. 

\subsection*{Kaluza-Klein
thermodynamics at finite density}

We consider a massless Dirac field in the space $R^d\times S^N$.
under such circumstances, we assume the thermodmamic potential $\Omega$
must be of the form
\begin{equation}
\Omega=\frac{1}{\beta}\left(\frac{R}{a}\right)^df(\mu a,
\beta/a)\equiv \frac{1}{\beta}\left(\frac{R}{a}\right)^df(x,
y)\,,  
\label{5.1}
\end{equation}
where $R$ is the scale factor of the flat $d$-dimensional space and $a$
is the radius of the $N$-dimensional hypersphere $S^N$.

The chemical potential is introduced for a fermion field in
$(d+N)$-dimensional space. In the Kaluza-Klein sense, we find
indefinitely many particles in $d$-dimensional space, however, the
particle number defined in this paper is to be conserved, because this
may be regarded as the charge asymmetry in the system.
 
Now let us derive several thermodynamic quantities using
Eq.~(\ref{5.1}). We obtain
\begin{eqnarray}
PV_dV_N&=& -\frac{1}{d}R\frac{\partial\Omega}{\partial R}=
-\frac{1}{\beta}\left(\frac{R}{a}\right)^df\,,\nonumber \\
 QV_dV_N&=& -\frac{1}{N}a\frac{\partial\Omega}{\partial a}=
-\frac{1}{N}\frac{1}{\beta}\left(\frac{R}{a}\right)^d
\left[-d\cdot f+x\frac{\partial
f}{\partial x}-y\frac{\partial f}{\partial y}\right]\,,\nonumber
\\
S&=&\beta^2\frac{\partial\Omega}{\partial\beta}=
\left(\frac{R}{a}\right)^d
\left[-f+y\frac{\partial f}{\partial y}\right]\,,\nonumber
\\
\mu{\cal N}&=&-\mu\frac{\partial\Omega}{\partial\mu}=
-\frac{1}{\beta}\left(\frac{R}{a}\right)^d
x\frac{\partial f}{\partial x}\,,\nonumber
\\
E&=&\Omega+\mu{\cal N}+\frac{1}{\beta}S=
\frac{1}{\beta}\left(\frac{R}{a}\right)^d
\left[-x\frac{\partial f}{\partial x}+y\frac{\partial f}{\partial
y}\right]\,, 
\label{5.2}
\end{eqnarray}
where $V_d$, $V_N$, $P$, $Q$, $S$, ${\cal N}$ and $E$ are, the volume of
$R^d$, the volume of $S^N$, the pressue for $R^d$, the pressure for
$S^N$, the entropy, the particle number and the internal energy,
respectively. We notice immediately,
\begin{equation}
 -\rho+d\cdot P+N\cdot Q=0\,,\quad\mbox{here}\quad
\rho=\frac{E}{V_dV_N}\,.      
\label{5.3}
\end{equation}
This relation has often been mentioned, and is equivalent to state
the thermal contribution to the stress tensor is traceless.

On the other hand, the conservation of the stress tensor gives
\begin{equation}
dE+PV_NdV_d+QV_ddV_N=TdS+\mu d{\cal N}=0\,.   
\label{5.4}
\end{equation}
It is natural to assume the entropy and the particle number are
conserved separately. This point will be discussed again in the next
section. 

\subsection*{Instability of compactification at finite density}

In order to gain stable compactifications of extra spaces, various
mechanisms are proposed. Candelas and Weinberg \cite{20} considered
quantum effects of matter fields in the case of the compact spaces,
$S^N$. Many other authors showed the extended versions of their model
and the presence of various sorts of instabilities.\cite{21,22}
Recently, Accetta and Kolb exhibited the finite temperature instability
for compactifications, and the critical temperature for the
instability.\cite{23} Here, we show the finite density instability,
which can be discussed almost parallel to the work by Accetta and Kolb.

At high density ($\beta \mu \gg 1$), the themodynamic potential for
Dirac fields can be written in a form
\begin{equation}
\Omega=\frac{V_d}{a^{d+1}}(C_N-C(\mu a)^{d+N+1})\,.          
\label{5.5}
\end{equation}
Note here $\Omega$ contains the Casimir stress energy which is
determined at one-loop level when the space-time dimensionality is odd.
The pressure for the internal space, $Q$, and the particle number ${\cal
N}$ are computed from (\ref{5.5}):
\begin{eqnarray}
Q&=& -\frac{1}{V_N}\frac{1}{N}a\frac{\partial\Omega}{\partial a}
\nonumber \\
&=&\frac{V_d}{V_N}\frac{1}{a^{d+1}}[(d+1)C_N+N\cdot
C(\mu a)^{d+N+1}]\,,\nonumber
\\ {\cal N}&=&-\frac{\partial\Omega}{\partial\mu}\nonumber \\
&=&\frac{V_d}{a^d}C(d+N+1)(\mu a)^{d+N}\,.          
\label{5.6}
\end{eqnarray}
We recognize the similarity to the high-temperature case.\cite{22} The
only difference is the exchange of $\mu$ and $T$. Therefore we can
immediately obtain the critical value for $\mu$:
\begin{equation}
\mu_{crit}\sim\frac{1}{a_0}\left[
\frac{C_N}{C}\frac{2(d-1)}{N}\left\{
\frac{d+1}{N}+1\right\}\right]^{1/(N+d+1)}  \,,
\label{5.7}
\end{equation}
where $a_0$ is the static radius of $S^N$. Using this, we conclude that
there are no stable compactifications when ${\cal N}/V_d>({\cal
N}/V_d)_{crit}$, where
\begin{equation}
\left(\frac{{\cal N}}{V_d}\right)_{crit}\sim C
\frac{d+N+1}{a_0^d}
\left[
\frac{C_N}{C}\frac{2(d-1)}{N}\left(
\frac{d+1}{N}+1\right)\right]^{(d+N)/(d+N+1)} 
\label{5.8}
\end{equation}
For example, we take $d=3$ and $N=7$. In this case,
\begin{eqnarray}
C_N&=&5.958744 \times 10^{-5} \times f\,,\nonumber\\
C&=&\frac{{\rm tr}{\bf 1}}{2}\frac{1}{(4\pi)^{(d+N)/2}}
\frac{2\pi^{(N+1)/2}/\Gamma\left(\frac{N+1}{2}\right)}{
\Gamma\left(\frac{d+N+2}{2}\right)(d+N+1)}\nonumber \\
&=&3.139894\times 10^{-7}\times f \,,                      
\label{5.9}
\end{eqnarray}
where $f$ denotes the number of Dirac fields. Then we obtain
\begin{equation}
\left(\frac{{\cal N}}{V_d}\right)_{crit}=\frac{f}{a_0^d}\times 3.35
\times 10^{-5}\,.          
\label{5.10}
\end{equation}
If we consider the initial size of 3-space is the same order as the
one of the compact space, Eq.~(\ref{5.10}) reads the severe constraint
to the charge asymmetry.

One can discuss more details (such as including semiclassical
instability \cite{22}), by repeating similar analyses of Accetta and
Kolb.\cite{23}

%%%%%%%%%%%%%%%%%%%%%%%%%%%%%%%%%%%%%%%%%%%%%%%%%%%%%%%
\section{Summary and discussion}
%%%%%%%%%%%%%%%%%%%%%%%%%%%%%%%%%%%%%%%%%%%%%%%%%%%%%%%
We gave the expressions for the thermodynamic potential of a Dirac
fermion gas with the chemical potential in the space $R^d\times S^N$. We
mainly pay our attention to the low-temperature case which leads to
the strolls degeneracy of fermi gas. We also gave some examples for
applications to a few aspects of Kaluza-Klein theories. We omitted
the detailed discussion about the application to the Kaluza-Klein
cosmology, which thus will be discussed in another occasion.

The treatment of finite density implicitly assumes the existence of
conserved $U(1)$ charge. Superstring theories imply the presence of the
gauge field as ``primary field,'' so we may expect some conserved
charges.
The chemical potentials accompanied with non-abelian charges are
also introduced by Haber and Weldon.\cite{11} The generalization to
the high-dimensional case will enable us to investigate the finite
density effect on the breakdown of primary gauge symmetries as well as
compactifications. It may also be interesting to consider the themal
property of the gauge field itself \cite{24} in higher dimensions.

On the other hand, it is also interesting to consider the chemical
potentials associated to each ``Kaluza-Klein charge,'' which are
induced after the compactiacation. This analysis will become necessary
when one studies the property of the pyrgons \cite{25} at finite
temperature and density. We leave this for future publications.
  
We need to consider the thermodynamic quantities for boson fields
for further study. There is the technical subtlety to calculate the
themodynamic potentials in relation to the bosonic field with the
chemical potential.\cite{26} But there are also attractive phenomena
such as the Bose-Einstein condensation. We will report on these
problems generalized to higher-dimensional case elsewhere.\cite{28}

%%%%%%%%%%%%%%%%%%%%%%%%%%%%%%%%%%%%%%%%%%%%%%%%%%%%%%%
\section*{Acknowledgments}
%%%%%%%%%%%%%%%%%%%%%%%%%%%%%%%%%%%%%%%%%%%%%%%%%%%%%%%
The author is very grateful to M.~Hosoda and T.~Hori for reading of the
manuscript and useful comments.

%%%%%%%%%%%%%%%%%%%%%%%%%%%%%%%%%%%%%%%%%%%%%%%%%%%%%%%
\section*{Appendix}
%%%%%%%%%%%%%%%%%%%%%%%%%%%%%%%%%%%%%%%%%%%%%%%%%%%%%%%
We explain the identity appeared in \S\S 2 and 3 in terms of theta
functions. First we observe:
\begin{equation}
\vartheta_3(v|\tau)=\exp(i\pi\tau n^2+i\pi 2nv)\, ,                
\label{A.1}
\end{equation}
and the well known relation:
\begin{equation}
\vartheta_3\left(\frac{v}{\tau}| -\frac{1}{\tau}\right)=(-
i\tau)^{1/2}\exp(i\pi v^2/\tau)\vartheta_3(v|\tau)\,.
\label{A.2}
\end{equation}
Using this, (\ref{2.8}) can be derived as follows:
\begin{eqnarray}
&&\sum_{n=-\infty}^\infty \exp\left\{ -t
\left(\frac{(2n+1)\pi}{\beta}+i\mu
\right)^2\right\}\nonumber \\
&& =\exp\left\{ -t
\left(i\mu+\frac{\pi}{\beta}
\right)^2\right\}\vartheta_3\left(\left.\frac{2i}{\beta}
\left(i\mu+\frac{\pi}{\beta}\right)t\right|\frac{4\pi}{\beta^2}it
\right)\nonumber
\\ 
&&
=\left(\frac{4\pi}{\beta^2}t\right)^{-1/2}\vartheta_3\left(\left.
\frac{\beta}{2\pi}
\left(i\mu+\frac{\pi}{\beta}\right)\right|\frac{\beta^2}{4\pi}
\frac{i}{t}
\right)\nonumber
\\&&
=\left(\frac{\beta^2}{4\pi
t}\right)^{1/2}\sum_{n=-\infty}^\infty
\exp\left(-\frac{\beta^2n^2}{4t}\right)\exp
\left\{i\beta\left(i\mu+\frac{\pi}{\beta}\right)n\right\}\nonumber
\\ &&
=\frac{\beta}{(4\pi)^{1/2}
}t^{-1/2}\left[1+2\sum_{n=1}^\infty (-1)^n
\exp\left(-\frac{\beta^2n^2}{4t}\right)\cosh(\mu\beta n)\right]\,.
\label{A.3}
\end{eqnarray}
 
Next, we show the derivation of (3.2). A similar expansion but for
bosons is given by Yoshimura.\cite{27} We consider the case $N$ is odd.
\begin{equation}
S_F\equiv\sum_{l=0}^\infty d_l\,\exp\{-(l+N/2)^2x\}\, ,
\quad\mbox{where}\quad
d_l=\frac{\Gamma(l+N)}{l!\,\Gamma(N)}\,.   
\label{A.4}
\end{equation}
We write the degeneracy $d_l$ as
\begin{eqnarray}
d_l&=&\frac{1}{\Gamma(N)}\left[
\left(l+\nu+\frac{1}{2}\right)^2-\left(\nu-\frac{1}{2}\right)^2
\right]\cdot\left[
\left(l+\nu+\frac{1}{2}\right)^2-\left(\nu-\frac{3}{2}\right)^2
\right]\nonumber \\
&&\cdots\left[
\left(l+\nu+\frac{1}{2}\right)^2-\left(\frac{1}{2}\right)^2
\right]\nonumber\\
&=&\sum_{m=0}^\nu C_{\nu m}\,\left(l+\nu+\frac{1}{2}\right)^{2m}\,,
\qquad
\nu=\frac{N-1}{2}\,.
\label{A.5}
\end{eqnarray}
$C_{\nu m}$ are independent of $N$, and the first two terms are given by
\begin{equation}
C_{\nu\nu}=\frac{1}{\Gamma(N)}\,,
\label{A.6}
\end{equation}
\begin{equation}
C_{\nu\,\nu-1}=-\frac{1}{\Gamma(N)}\sum_{l=1}^\nu
\left(l-\frac{1}{2}\right)^2
=-\frac{1}{24}\frac{1}{\Gamma(N)}N(N-1)(N-2)\,. 
\label{A.7}
\end{equation}
Since the sum over $m$ vanishes when $l=-1, -2,\cdots, -\nu$, we carry
out the summation over $l$ from $l=-\nu$ to $\infty$. Then, substituting
(\ref{A.5}) into (\ref{A.4}), we obtain
\begin{eqnarray}
S_F&=&\sum_{m=0}^\nu C_{\nu m} (-1)^m\frac{d^m}{dx^m}
\sum_{l=-\nu}^\infty\exp\left[-\left(
l+\nu+\frac{1}{2}\right)^2x\right]\nonumber \\
&=&\sum_{m=0}^\nu C_{\nu m} (-1)^m\frac{d^m}{dx^m}
\left[\frac{1}{2}\sqrt{\frac{\pi}{x}}+\sqrt{\frac{\pi}{x}}
\sum_{n=1}^\infty (-1)^n\exp\left(-
\frac{\pi^2n^2}{x}\right)\right]\,.
\label{A.8}
\end{eqnarray}
Here, we have used the identity derived from (\ref{A.2})
\begin{eqnarray}
\sum_{l=-\infty}^\infty \exp\left[-\left(
l+\frac{1}{2}\right)^2 x\right]&=&\exp\left(- \frac{x}{4}\right)
\vartheta_3\left(\left.\frac{ix}{2\pi}\right|\frac{ix}{\pi}\right)
\nonumber \\
&=&\sqrt{\frac{\pi}{x}}
\vartheta_3\left(\left.\frac{1}{2}\right|i\frac{\pi}{x}\right)
\nonumber \\
&=&\sqrt{\frac{\pi}{x}}\sum_{n=-\infty}^\infty (-1)^n
\exp\left(-\frac{\pi^2n^2}{x}\right)\,.
\label{A.9}
\end{eqnarray}
From (\ref{A.8}) and (\ref{A.5}, \ref{A.6}, \ref{A.7}), we find the
asymptotic form of $S_F$,
\begin{equation}
S_F\stackrel{x\rightarrow
0}{\longrightarrow}\frac{1}{2}
\frac{\Gamma(N/2)}{\Gamma(N)}x^{-N/2}\left(1-\frac{1}{12}N(N-1)x
+O(x^2)\right)\,.    
\label{A.10}
\end{equation}

%%%%%%%%%%%%%%%%%%%%%%%%%%%%%%%%%%%%%%%%%%%%%% 

%%%%%%%%%%%%%%%%%%%%%%%%%%%%%%%%%%%%%%%%%%%%%
\end{document}